\documentclass[paper]{JHEP}
\usepackage{epsfig}

\newcommand{\beq}{\begin{equation}}
\newcommand{\eeq}{\end{equation}}
\newcommand{\beqa}{\begin{eqnarray}}
\newcommand{\eeqa}{\end{eqnarray}}
\newcommand{\eq}[1]{Eq.\ (\ref{#1})}

\def\gsim{\ \rlap{\raise 3pt \hbox{$>$}}{\lower 3pt \hbox{$\sim$}}\ }
\def\lsim{\ \rlap{\raise 3pt \hbox{$<$}}{\lower 3pt \hbox{$\sim$}}\ }

\newcommand{\order}[1]{${\cal O}(#1)$}

\newcommand{\cf}{{\it cf.\ }}

\newcommand{\as}{\alpha_s}
\newcommand{\ieps}{i\varepsilon}
\newcommand{\kvec}{{\mbox{\bf k}}}
\newcommand{\skvec}{{\mbox{\scriptsize\bf k}}}
\newcommand{\qvec}{{\mbox{\bf q}}}

\newcommand{\ket}[1]{\vert{#1}\rangle}
\newcommand{\bra}[1]{\langle{#1}\vert}

\newcommand{\den}[1]{(#1+\ieps)^2}
\newcommand{\la}{\lambda}
\newcommand{\Pit}{\tilde\Pi}
\newcommand{\at}{\tilde\alpha_V}
\newcommand{\bt}{\tilde\beta_0}

\title{The static PQCD potential with modified boundary conditions}

\author{Paul Hoyer\footnote{On leave of absence from the Department 
of Physics, University of Helsinki.} \\ 
Nordita, Blegdamsvej 17, DK--2100 Copenhagen,
Denmark\\ E-mail: \email{hoyer@nordita.dk}}

\author{Johan Rathsman\\
CERN, TH-Division, CH--1211 Geneva 23, Switzerland\\
E-mail: \email{Johan.Rathsman@cern.ch}}

\preprint{ CERN-TH/2000-334\\
   NORDITA-2000-96 HE\\
   Revised 4 May, 2001\\  
   \hepph{0011209}}

\abstract{
We calculate the potential between two static quarks in QCD using modified
boundary conditions for the perturbative expansion. Through a change of the
Feynman $\ieps$ prescription we effectively add a ``sea'' of gluons to the
asymptotic states with energies below a given scale $\Lambda$.
We find that the standard result for the static potential gets
corrections of order $\Lambda^2/Q^2$ both at small and large
momentum transfers $Q^2$.
After resummation of the infrared sensitive corrections we find that
the running coupling $\as(Q^2)$ freezes in the infrared and that the
exchanged gluon gets an effective tachyonic mass.
We verify that identical results are obtained in the Coulomb and Feynman
gauges.}


\begin{document}

\section{Introduction}

In perturbative QCD (PQCD) calculations of S-matrix amplitudes quarks and
gluons are assumed to form free asymptotic states at the initial and final
times, $t\to \pm\infty$. It is recognized that this is at variance with
observations -- partons actually bind to form colour singlet hadrons
which are the true asymptotic states. Consequently, the applications of
perturbation theory are restricted to so-called infrared safe observables in
processes characterized by a large momentum scale $Q$. All predictions are
subject to power corrections
$(\Lambda/Q)^n$, where $\Lambda \sim 200$ MeV is the fundamental QCD scale.

It has been noted \cite{Dokshitzer} that PQCD predictions can nevertheless be
successfully extrapolated to low scales $Q\sim \Lambda$, assuming that the
$Q$-dependence of the running coupling $\as(Q^2)$ ``freezes'' at
a hadronic scale of order $\Lambda$. Confinement appears to change
momentum distributions only in a mild way, with PQCD distributions of partons
being reflected in those observed for hadrons. This motivates us
to study whether PQCD can be modified so that its use can be extended to low
$Q^2$ without having to introduce the freezing effects ``by hand''.

Formally, there is considerable freedom in making a perturbative expansion.
The standard arguments justifying an expansion, namely
\begin{itemize}
\item The initial and final times are taken to infinity along a ray slightly
tilted {\it wrt.} the real axis, and
\item The asymptotic configurations have a non-vanishing overlap with the true
ground state of the theory
\end{itemize}
allow many choices of {\it in-} and {\it out-}states. The
existence of an overlap with the true ground state is in practice an
assumption,
even in the case of standard PQCD where the asymptotic states are taken to be
the empty ``perturbative vacuum''. Considering the central importance of
perturbation theory in applications of field theory it seems desirable to
explore the properties of expansions with different asymptotic states.

Here we will study the effect of adding gluons to the perturbative vacuum. It
is natural to consider this since the true QCD ground state is believed to be
a condensate of gluons. Conceivably, the background gluons may mimic the
properties of the true gluon condensate sufficiently to make the perturbative
expansion express some of the confinement physics already at low orders. In any
case, the above formal arguments justifying such a modified perturbative
expansion are as compelling as those of standard PQCD.

The  specific modification of the asymptotic state we consider has been called
the ``Perturbative Gluon Condensate'' \cite{Hoyer}. Background gluons with
energies smaller than a given scale $\Lambda$ are introduced by modifying the
Feynman $\ieps$ prescription of the gluon propagator in the following way:
\begin{eqnarray}
\frac{1}{k^2+i\varepsilon}& \to & \frac{1}{\den{k}}  \equiv
\frac{1}{k^2+i\varepsilon} + \frac{i\pi}{2|\kvec|}
\left[\delta(k_0-|\kvec|)+\delta(k_0+|\kvec|) \right] \Theta(\Lambda-|\kvec|)
   \label{eq:modiepsilon} \\
&& = \frac{\Theta(|\kvec|-\Lambda)}{k^2+i\varepsilon} +
\frac{1}{2}
\left[\frac{1}{(k_0-\ieps)^2-\kvec^2} + \frac{1}{(k_0+\ieps)^2-\kvec^2}\right]
\Theta(\Lambda-|\kvec|) \, ,
\nonumber
\end{eqnarray}
where $1/({k^2+i\varepsilon})$ denotes the ordinary Feynman 
$\ieps$ prescription and $1/\den{k}$ denotes the modified one. As was shown
\cite{Hoyer} for scalar fields, a perturbative calculation of any Green
function $G$ using the modified propagator (\ref{eq:modiepsilon}) is equivalent
to a superposition of standard calculations using Feynman propagators with
gluons added to the asymptotic states, schematically
\begin{eqnarray}\label{eq:superposition}
\bra{0} G \ket{0} &\to&
\left( \prod_{|\skvec|<\Lambda}\, \sum_{n_\skvec =0}^{\infty} c_{n_\skvec}
\right)\, \bra{\,\prod_\skvec (g_\skvec)^{n_\skvec}}\, G\,
\ket{\,\prod_\skvec(g_\skvec)^{n_\skvec}} \, .
\end{eqnarray}
Here the $n_\skvec=0$ term corresponds to the
unmodified expansion, the $c_{n_\skvec}$
are known constants and the sum is over on-shell gluons $g_\skvec$ of
momentum $\kvec$ and energy $|\kvec|<\Lambda$. We will show here that
gauge invariance is maintained when both gluon and ghost propagators are
modified according to \eq{eq:modiepsilon}.

Physically, the modified asymptotic states imply scattering off the
``background'' gluons which prevents the creation of gluons with
$|\kvec| < \Lambda$. Technically this can be seen from the sign change
(\ref{eq:modiepsilon}) of $\ieps$ in the free gluon propagator which removes
pinches between positive and negative energy poles in loop integrals. For a
fermion propagator, such a change of $\ieps$ would be equivalent to filling all
fermion (or antifermion) levels up to a Fermi momentum $\Lambda$, and
consequently preventing fermion pair production in accordance with the Pauli
exclusion principle. We are motivated to study the analogous modification of
the gluon propagator as a way of avoiding the production of soft gluons in
perturbation theory. Since we effectively superpose calculations with different
numbers of background gluons as indicated in \eq{eq:superposition}, we need not
specify the wave function of such a ``Dirac gluon sea'' (\cf \cite{NN}). We
shall refer to the physics based on the modified gluon propagator
(\ref{eq:modiepsilon}), with the standard Feynman $\ieps$ prescription for
quark propagators, as ``Perturbative Gluon Condensate Dynamics'', or PGCD.
Formally, the PGCD expansion appears as justified as ordinary PQCD.

The introduction of a fixed momentum scale $\Lambda$ in the PGCD propagator
(\ref{eq:modiepsilon}) seems to break Lorentz invariance. The perturbative
expansion of the amplitude for a given process will depend on the reference
frame, since the  scale $\Lambda$ is frame independent. Formally the series
sums to the same (Lorentz covariant) result in any frame, but the rate of
convergence is frame dependent. The situation is in this sense analogous to the
well-known freedom of choice in the renormalization scale. Physical arguments
must be used to choose an optimal frame for each process. This is in fact
commonly done in hadron phenomenology. The non-relativistic quark
model describes hadrons in their rest frames, whereas the parton model is
formulated in the infinite momentum frame.

We should emphasize that the boost properties of bounds states are in general
extremely complicated \cite{dirac}. In QED, positronium wave functions and
energy levels are nearly always evaluated in the rest frame, and most
efficiently using non-covariant methods such as NRQED \cite{nrqed}. Not even
general features such as the Lorentz contraction of QED bound states have (to
our knowledge) been explicitly demonstrated in perturbation theory. In QCD we
face the extra challenge that the gluon condensate ground state is boost
invariant: the gluons carry momenta of \order{\Lambda_{QCD}} in any frame. This
feature can clearly not be described using perturbation theory -- the best we
can do is to approximate the true ground state with background gluons whose
momenta are the same in any frame, as in \eq{eq:modiepsilon}.

In this paper we consider the effects of  PGCD on the static quark potential.
This implies an automatic frame choice since the static potential is defined
only in the ``rest frame'' of the static sources. We shall not further discuss
the important and non-trivial question of Lorentz invariance. The question of
frame choice for a general process is beyond the scope of this paper.

According to the Kinoshita-Lee-Nauenberg (KLN) theorem~\cite{KLN} all
infrared singularites cancel if one sums over incoming and outgoing states
that are degenerate in energy. Our procedure of adding soft gluons to the {\it
in-} and {\it out-}states introduces a similar smearing of the physical
observables. It has in fact been
shown~\cite{Akhoury:1996cj,Akhoury:1997pb,Akhoury:1998ks} that the
``KLN-cancellations'' can be accounted for using a similar modification of the
$\ieps$ prescription as the one we study here. As discussed
in~\cite{Akhoury:1998ks} the effects of the KLN-cancellation can be thought of
as a ``KLN vaccum'' and the non-vanishing interactions with the vacuum as
``perturbative condensates''. Thus the physical picture appears similar to the
PGCD. The KLN-cancellations are valid in any field theory irrespective of
whether there is confinement or not, and the energy-resolution (corresponding
to $\Lambda$) can be arbitrarily small. In our interpretation the scale of soft
gluons is a physical feature related to the ground state of QCD.

The purpose of this paper is two-fold. On the one hand we want to investigate
whether the PGCD boundary conditions give a perturbative expansion
which captures some of the physics of QCD at long distances, while leaving
unchanged standard perturbative results at short distances. As a
first test case
we calculate the QCD potential between two  static colour sources in a colour
singlet state~\cite{Susskind:1976pi}. We compare the ultraviolet and infrared
properties of the static PGCD potential with results obtained using
ordinary PQCD. The second purpose of this paper is to check explicitly that the
perturbative gluon condensate framework is gauge invariant. Hence we do
the calculation both in a physical and in a covariant gauge, namely the Coulomb
and Feynman gauges.

\section{Calculation of static potential}

The QCD potential $V(Q^2)$ between two static colour sources can be defined in
a gauge invariant way from a Wilson loop~\cite{Susskind:1976pi}.
At lowest order the potential is just given by one-gluon exchange,
$V(Q^2)= - C_F {g^2}/{Q^2}$, where $g^2$ is the strong coupling and
$q^2=-Q^2=-\qvec^2$ is the squared momentum transfer
which is purely space-like in the static approximation, {\it i.e.}\ $q_0=0$.
The PGCD $i\varepsilon$ prescription does not change this lowest order result
since the coupling of the background gluons
to a source with large mass $M$ is suppressed by $\Lambda/M$.
At higher orders the fixed coupling $g^2$ is  replaced by the
running coupling after renormalization. Including all higher-order corrections
in the running coupling gives an effective charge $\alpha_V(Q^2)$ defined by
\begin{eqnarray}
V(Q^2) \equiv - 4\pi C_F \frac{\alpha_V(Q^2)}{Q^2} \,
\end{eqnarray}
where $C_F=(N_C^2-1)/2N_C=4/3$ for QCD.
In the following we will calculate $\alpha_V(Q^2)$ to one-loop
order using the PGCD $i\varepsilon$ prescription (\ref{eq:modiepsilon}). For
convenience we define the one-loop correction
$\hat{\Pi}(Q,Q_0,\Lambda)$ so that
the leading order result is factored out,
\begin{eqnarray}
\alpha_V(Q^2) =   
   \alpha_V(Q_0^2)\left[1+\hat{\Pi}(Q,Q_0,\Lambda) + \cdots \right]  \, , 
\end{eqnarray} 
where $Q_0$ is the renormalization point, 
{\it i.e.,}\ $\hat{\Pi}(Q_0,Q_0,\Lambda)=0$.

\subsection{Coulomb gauge}
Coulomb gauge is the most natural gauge for calculating the static potential
\cite{khriplovich}, although the Feynman rules are not as simple
as in a covariant gauge such as Feynman gauge. Here we will use the Feynman
rules of Coulomb gauge given by Feinberg~\cite{Feinberg:1978rc}.
The diagrams contributing to the static potential at one-loop order in
Coulomb gauge are shown in Fig.~\ref{fig:coulomb}. For clarity we do not
include the contribution from light quarks, which is the same as in
standard PQCD.

\FIGURE[t]{\epsfig{figure=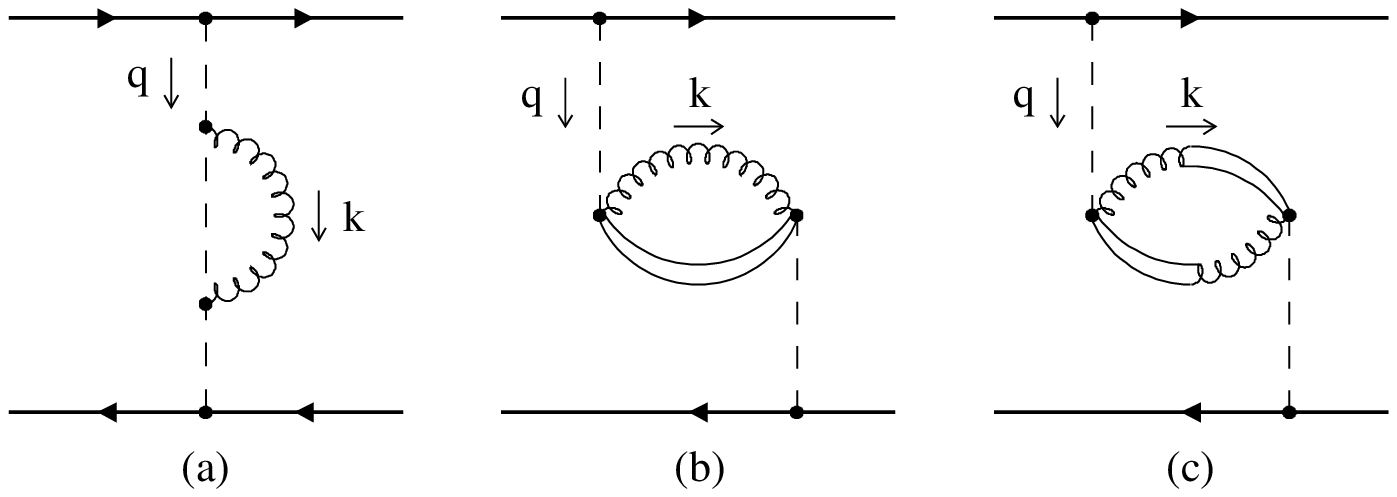,width=12cm}
\caption{One-loop diagrams contributing to the static potential
in Coulomb gauge. The thick lines represent the static quarks,
and the dashed lines the instantaneous Coulomb propagators. The
curly and double lines represent the $A$ and $E$-field propagators,
respectively. There is also a mixed $A$ and $E$-field propagator
which appears in (c). }
\label{fig:coulomb}}

Using dimensional regularization the contribution to
the unrenormalized one-loop correction $\Pi$ from
the Coulomb self-energy diagram shown in Fig.~\ref{fig:coulomb}(a) is
\begin{eqnarray}
\Pi_{a} &=& 3ig^2C_A \mu^{4-n} \int \frac{d^nk}{(2\pi)^4}
\qvec_i\qvec_j\left(\delta_{ij}-\frac{\kvec_i\kvec_j}{\kvec^2}\right)
\frac{1}{\qvec^2(\qvec-\kvec)^2}
\frac{1}{\den{k}}
\end{eqnarray}
where $C_A=N_C=3$,
$n$ is the number of dimensions ($n<4$), $\mu$ is the arbitrary
dimensional regularization scale,
the subscripts $i,j$ denote the space components ($i,j=1,2,3$), and
the $i\varepsilon$ prescription for the transverse gluon propagator is given in
Eq.~(\ref{eq:modiepsilon}).
We have written the integrand in 4 dimensions since
we will not be interested in constant contributions to $\Pi$.
This corresponds to a specific choice of renormalization scheme.
The $k_0$-integral is conveniently done in Minkowski space using ordinary
residue calculus, and vanishes for $|\kvec|< \Lambda$ since the poles at
$k_0=\pm |\kvec|$ are then on the same side of the real axis. The result is
symmetric under $\kvec \leftrightarrow \qvec-\kvec$ and can be expressed as
\begin{eqnarray}
\Pi_{a} &=& 3g^2C_A \mu^{4-n} \int \frac{d^{n-1}k}{(2\pi)^3}
\left(1-\frac{(\kvec\cdot\qvec)^2}{\kvec^2\qvec^2}\right)
\nonumber \\ &&
\times\frac{1}{(\qvec-\kvec)^2}
\frac{1}{2}
\left[\frac{\Theta(|\kvec|-\Lambda)}{2|\kvec|} +
\frac{\Theta(|\qvec-\kvec|-\Lambda)}{2|\qvec-\kvec|} \right]
\end{eqnarray}
where the $\Theta$-functions reflect
the modified $i\varepsilon$ prescription.

To simplify the remaining integrations it is convenient to choose
\begin{eqnarray}
x  =  \frac{|\kvec|}{Q} \ ,\hspace{2cm}
y  =  \frac{|\qvec-\kvec|}{Q} \label{xydef}
\end{eqnarray}
as new integration variables, with measure
\begin{eqnarray} \label{jacobian}
\int d^{n-1}k & = & \int_0^\infty dx  \,  \int_{|x-1|}^{x+1} dy \,
\int_0^{2\pi} d\varphi
\, Q^{n-1} x^{n-3} y \,
\end{eqnarray}
where we have again dropped terms proportional to $(n-4)$ in the
angular integral.
The integration over the azimuthal angle $\varphi$ gives
$2\pi$ and the remaining integral becomes
\begin{eqnarray}
\Pi_a &=& 3C_A\frac{g^2}{4\pi^2} \left(\frac{\mu}{Q}\right)^{4-n}
\int_0^\infty dx \, x^{n-4} \int_{|x-1|}^{x+1} dy
\frac{-x^4-y^4+2x^2y^2+2x^2+2y^2-1}{16x^2y^2}
\nonumber \\ &&
\times\left[y\Theta(x-\la) + x\Theta(y-\la) \right] ,
\label{eq:Pia}
\end{eqnarray}
where $\la=\Lambda/Q$. Before evaluating the integral we shall add the
contributions from the remaining diagrams to the integrand.

According to the rules given by Feinberg~\cite{Feinberg:1978rc},
the contribution to $\Pi$ from the sum of the vacuum-polarization diagrams
in Fig.~\ref{fig:coulomb}(b) and (c) is
\begin{eqnarray}
\Pi_{b+c} &=& ig^2C_A \mu^{4-n} \int \frac{d^nk}{(2\pi)^4}
\left(\delta_{ij}-\frac{(\qvec-\kvec)_i(\qvec-\kvec)_j}
{(\qvec-\kvec)^2}\right)
\left(\delta_{ij}-\frac{\kvec_i\kvec_j}{\kvec^2}\right)
\nonumber \\ &&
\times\frac{k_0^2+\frac{1}{2}[\kvec^2+(\qvec-\kvec)^2]}
{\qvec^2 \den{k}\den{q-k}} \;.
\end{eqnarray}
After integrating over $k_0$ and $\varphi$ and using
(\ref{xydef}) and (\ref{jacobian}) this becomes
\begin{eqnarray}
\Pi_{b+c} &=& C_A\frac{g^2}{4\pi^2} \left(\frac{\mu}{Q}\right)^{4-n}
\int_0^\infty dx \, x^{n-4} \int_{|x-1|}^{x+1} dy
\frac{x^4+y^4+6x^2y^2-2x^2-2y^2+1}{16x^2y^2}
\nonumber \\ &&
\times\left[y\frac{3x^2+y^2}{x^2-y^2}\Theta(x-\la) +
x\frac{3y^2+x^2}{y^2-x^2}\Theta(y-\la) \right] \, .
\label{eq:Pibc}
\end{eqnarray}

Adding the Coulomb self-energy and vacuum-polarization contributions of
Eqs.~(\ref{eq:Pia}) and (\ref{eq:Pibc}) gives
\begin{eqnarray}
\Pi &=& C_A\frac{g^2}{4\pi^2} \left(\frac{\mu}{Q}\right)^{4-n}
\int_0^\infty dx \, x^{n-4} \int_{|x-1|}^{x+1} dy
\left[\frac{7x^4+y^4-2x^2-2y^2+1}{4x^2(x^2-y^2)}y\Theta(x-\la)
\right. \nonumber \\ && + \left.
\frac{7y^4+x^4-2y^2-2x^2+1}{4y^2(y^2-x^2)}x\Theta(y-\la) \right] \, ,
\label{CPi}
\end{eqnarray}
Note that the apparent pole at $x=y$ cancels between the two
terms in the integrand.
Doing the integrals we find the  result for the unrenormalized one-loop
correction to the static potential,
\begin{eqnarray}\label{eq:Piunren}
\Pi(Q,\mu,\Lambda) &=& C_A\frac{g^2}{4\pi^2} \left[
\frac{11}{6}\ln\frac{2\mu}{Q(2\la+1)} + \frac{11}{6}\frac{1}{4-n} +
\frac{4}{3}\la^2+C \right. \nonumber \\ && + \left.
\frac{(2\la-1)(4\la^3+2\la^2-5\la+3)}{12\la}
\ln\frac{2\la+1}{|2\la-1|} \right] \, , 
\end{eqnarray} 
where $\la=\Lambda/Q$ and C is a renormalization-scheme-dependent constant.
This is the main result of our calculation. Before analysing it in
more detail we check that we get the same result if we do the
calculation in Feynman gauge. This will at the same time constitute a
non-trivial verification of the gauge invariance of the PGCD
$i\varepsilon$ prescription. 

\subsection{Feynman gauge}

\FIGURE[t]{\epsfig{figure=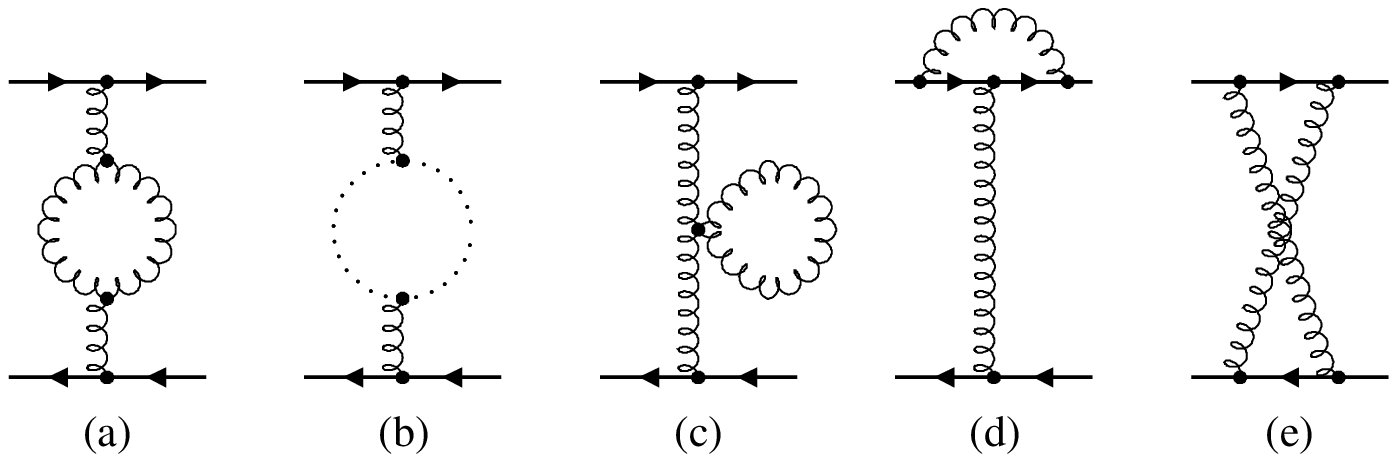,width=12cm} \caption{One-loop
diagrams contributing to the static potential in Feynman gauge. The
thick lines represent static quark propagators, the curly lines gluon
propagators and the dotted lines ghost propagators. For clarity the
vertices have been marked with dots.} \label{fig:feynman}} 

The diagrams which contribute to the static potential in Feynman gauge at
one-loop order are shown in Fig.~\ref{fig:feynman}. In addition to
the gluon propagator corrections of Fig.~\ref{fig:feynman}(a-c) there
is also the vertex correction of Fig.~\ref{fig:feynman}(d), which has
a non-Abelian contribution that does not cancel against the quark
wave-function renormalization, as well as the crossed box diagram of
Fig.~\ref{fig:feynman}(e), which has a non-Abelian part that is not
part of the iteration of the one-gluon exchange. In a general
covariant gauge a diagram with a three-gluon vertex also contributes,
but it vanishes in Feynman gauge. For more details on the diagrams
that contribute in Feynman gauge and how the iteration of the
one-gluon exchange works we refer to Fischler~\cite{Fischler:1977yf}.

Note that we have included the diagram with a four-gluon-vertex shown
in Fig.~\ref{fig:feynman}(c). In dimensional regularization this
diagram does not contribute to the logarithmic UV-divergence, only to
a quadratic divergence which normally cancels against the other two
gluon propagator corrections. However, since we are modifying the
$i\varepsilon$ prescription these cancellations are no longer
guaranteed and therefore we
include all diagrams.

We again use dimensional regularization and calculate the
integrands of all diagrams in 4 dimensions since
we are not interested in constant contributions to the final expression. The
result after performing the numerator and colour algebra is
\begin{eqnarray}
\Pi &=& \frac{ig^2}{2\qvec^2}C_A \mu^{4-n}
\int \frac{d^nk}{(2\pi)^4}
\left[ \frac{k^2+(k+q)^2 +4q^2+10k_0^2}{\den{k}\den{k+q}}
-\frac{2k_0^2}{\den{k}\den{k+q}}\right. \nonumber \\ && \left.
-\frac{6}{\den{k}} -\frac{2q^2}{(k_0+i\varepsilon)^2 \den{k}}
+\frac{q^4}{(k_0+i\varepsilon)^2 \den{k}\den{k+q}} \right] \, ,
\end{eqnarray}
where each term corresponds to a specific diagram in Fig.~\ref{fig:feynman}.
The $1/(k_0+i\varepsilon)$ factor in
the vertex correction and box diagrams comes from the static quark
propagator and is not to be confused with
the PGCD prescription (\ref{eq:modiepsilon}).
Doing the integrals over $k_0$ and $\varphi$
and making the variable substitutions $x = |\kvec|/Q$ and
$y=|\kvec+\qvec|/Q $ we are left with
\begin{eqnarray}
\Pi &=&  C_A\frac{g^2}{4\pi^2} \left(\frac{\mu}{Q}\right)^{4-n}
\int_0^\infty dx \, x^{n-4} \int_{|x-1|}^{x+1} dy
\left[\frac{6x^4+2x^2y^2-3x^2-y^2+1}{4x^2(x^2-y^2)}y\Theta(x-\la)
\right. \nonumber \\ && + \left.
\frac{6y^4+2y^2x^2-3y^2-x^2+1}{4y^2(y^2-x^2)}x\Theta(y-\la) \right]
\end{eqnarray}
Even though the integrand is different from the one of \eq{CPi} obtained in
Coulomb gauge, the final result after the integrals are done only differs
from \eq{eq:Piunren} by
a renormalization-scheme-dependent constant. There is thus full agreement
between the two calculations.

\section{Discussion of result}
Our renormalized result
for the one-loop contribution to the
static potential using the PGCD $i\varepsilon$ prescription is
\begin{eqnarray}\label{eq:Piren}
\hat{\Pi}(Q,Q_0,\Lambda) &=& C_A\frac{\alpha_V(Q_0^2)}{\pi} \left[
\frac{11}{6}\ln\frac{2\Lambda+Q_0}{2\Lambda+Q}
+
\frac{4}{3}\frac{\Lambda^2}{Q^2} +C
\right. \nonumber \\ && + \left.
\frac{2\Lambda-Q}{12\Lambda}
\left(4\frac{\Lambda^3}{Q^3}
      +2\frac{\Lambda^2}{Q^2}
      -5\frac{\Lambda}{Q}+3\right) 
\ln\frac{2\Lambda+Q}{|2\Lambda-Q|}
\right] \, , 
\end{eqnarray} 
which is obtained from Eq.~(\ref{eq:Piunren}) by making a subtraction
at $Q=Q_0$. The constant $C$ is thus determined by the condition
$\hat{\Pi}(Q_0,Q_0,\Lambda) = 0$.

A basic control of the validity of \eq{eq:Piren} is
that it agrees with the standard PQCD result in the $Q \to\infty$ limit.
For $\Lambda/Q \to 0$ we get
\begin{eqnarray}\label{eq:UVPi}
\left.\hat{\Pi}(Q,Q_0,\Lambda)\right|_{\Lambda/Q \to 0}
&=& C_A\frac{\alpha_V(Q_0^2)}{\pi} \left[
\frac{11}{6}\ln\frac{Q_0}{Q}
-\frac{\Lambda^2}{3Q^2} + \frac{\Lambda^2}{3Q_0^2}
+{\cal O}\left(\frac{\Lambda^4}{Q^4}-\frac{\Lambda^4}{Q_0^4}\right)
\right] \, .
\end{eqnarray}
Thus the ordinary asymptotic freedom~\cite{Gross:1973id,Politzer:1973fx}
result is retained with power-corrections $\Lambda^2/Q^2$.
Returning to the complete expression (\ref{eq:Piren})
we also note that $\hat{\Pi}$ is well defined for
all finite $Q/\Lambda$, including
$Q=2\Lambda$.
More precisely, $\hat{\Pi}$ is
continuous at $Q=2\Lambda$ but the derivative
$d\hat{\Pi}/d\ln{Q}$ has an (integrable) singularity at that point.

The leading power-correction in Eq.~(\ref{eq:UVPi}) scales as $\Lambda^2/Q^2$.
By contrast, in the operator product expansion one expects a $\Lambda^4/Q^4$
scaling behaviour (see~\cite{Balitsky:1985iw} for a phenomenological
calculation
and~\cite{Chetyrkin:1999yr} for a related discussion). In this sense our result
is more similar to the gluon propagator in   the manifestly gauge dependent
$<A_{\mu}^2>$ gluon condensate~\cite{Lavelle:1988eg}, which has been argued
recently to have a  possible physical
meaning~\cite{Gubarev:2001eu,Gubarev:2001nz}. A $\Lambda^2/Q^2$ scaling of the
power-corrections to the potential  in momentum space was also found in an
infrared renormalon analysis  by Beneke~\cite{Beneke:1998rk}. In this context
we  note that it is not possible to make direct comparisons of results obtained
for large $Q$ with calculations made in coordinate space since the Fourier
transform  from momentum space to coordinate space involves an integral over
all momenta $Q$.

The sign of the power-correction  in  Eq.~(\ref{eq:UVPi}) decreases the running
of the coupling since the sign of $(\Lambda^2/Q_0^2-\Lambda^2/Q^2)$ is opposite
to that of $\ln(Q_0/Q)$. An  opposite behaviour, namely infrared sensitive
short-distance corrections which lead to a confining potential were  found
recently~\cite{Akhoury:1998by}.  Since this calculation was made in coordinate
space the results cannot be directly compared as explained above. We also note
that the infrared renormalon analysis cannot predict the sign of the
power-correction, only its scaling~\cite{Beneke:1998rk}. To see whether the
negative sign of the power-correction found in  Eq.~(\ref{eq:UVPi}) gives a
freezing coupling or a confining potential we have to study the small $Q$
behaviour of Eq.~(\ref{eq:Piren}) since a  possible fixed point for the
evolution equation is at $Q=0$. 

Expanding our result (\ref{eq:Piren}) in the limit $Q/\Lambda \to 0$ we find 
\begin{eqnarray} \label{IRPi}
\left.\hat{\Pi}(Q,Q_0,\Lambda)\right|_{Q/\Lambda \to 0} 
&=& C_A\frac{\alpha_V(Q_0^2)}{\pi} 
\left[ C(Q_0,\Lambda) +
2\frac{\Lambda^2}{Q^2} 
+{\cal O}\left(\frac{Q^2}{\Lambda^2}\right)
\right] \, , 
\end{eqnarray} 
where $C(Q_0,\Lambda)$ is a constant. 
We note several interesting aspects of this. First of all we see that
the only infrared-sensitive term is of the form $\Lambda^2/Q^2$; all
other terms are either constant or vanish in the limit $Q/\Lambda
\to 0$. Especially there is no logarithmic $Q$-dependence in this
limit, in other words there is no logarithmic running of the coupling
for small $Q/\Lambda$. (This can also easily be seen directly from
Eq.~(\ref{eq:Piren}).) Another interesting property of (\ref{IRPi})
is that the sign of the quadratic infrared divergence $\Lambda^2/Q^2$
is opposite to the one found in Eq.~(\ref{eq:UVPi}) and thus
corresponds to a linear confining potential when Fourier-transformed
to coordinate space. 

On the other hand, the $\Lambda^2/Q^2$ term signals a possible breakdown of
our  expression for the static potential at small $Q^2$. A closer analysis of
its origin  in the Feynman gauge calculation shows that it arises in the
diagrams with insertions in the single gluon propagator shown in
Fig.~\ref{fig:feynman}(a-c).  Power counting shows that this is also true in a
general covariant gauge. Since these insertions can be iterated the
corresponding corrections should be resummed as a geometric series,
\begin{eqnarray}
V(Q^2) & = &
- 4\pi C_F\frac{\alpha_V(Q_0^2)}{Q^2}
    \left[1+\hat{\Pi}(Q,Q_0,\Lambda)  + \cdots
	\right]
\nonumber \\ & = &
- 4\pi C_F\frac{\alpha_V(Q_0^2)}{Q^2}
    \left[1+\Pit(Q,Q_0,\Lambda)
    + 2C_A\frac{\alpha_V(Q_0^2)}{\pi}\frac{\Lambda^2}{Q^2}
    \right. \nonumber \\  &&  \left.
   \hspace{3cm} +
\left(2C_A\frac{\alpha_V(Q_0^2)}{\pi}\frac{\Lambda^2}{Q^2}\right)^2
    + \cdots
	\right]
\nonumber \\ & = &
- 4\pi C_F\frac{\alpha_V(Q_0^2)}{Q^2-\nu^2}
    \left[1+\Pit(Q,Q_0,\Lambda) + \cdots \right]
  \end{eqnarray}
where $\nu^2=2C_A\alpha_V(Q_0^2)\Lambda^2/\pi$ is a tachyonic effective
gluon mass squared, $m_{g,\mbox{\tiny eff}}^2=-\nu^2$
and $\Pit$ is the remainder of $\hat{\Pi}$ after
subtracting the quadratically divergent contribution $\nu^2/Q^2$.
At higher orders in $g^2$ there will be other contributions
$\propto \Lambda^2/Q^2$ which will make the effective mass scale dependent.
We note that according to Chetyrkin, Narison and
Zakharov~\cite{Chetyrkin:1999yr} the phenomenology of a tachyonic gluon mass
is quite successful and suggests $\nu^2\sim0.5$~GeV$^2$.
More generally, the tachyonic pole indicates a qualitative change with
decreasing $Q^2$ in the physics described by PGCD. The implications of this
are beyond the scope of the present paper and require further study.

The remaining one-loop correction $\Pit$ can be absorbed
into a modified running coupling $\at(Q^2,\Lambda^2)$,
allowing our result to be expressed as
\begin{eqnarray}\label{eq:Vres}
V(Q^2) & = &
- 4\pi C_F \frac{\at(Q^2,\Lambda^2)}{Q^2-\nu^2} \, .
\end{eqnarray}
Since $\Pit$ goes to a
constant as $Q/\Lambda \to 0$ the modified coupling
$\at(Q^2,\Lambda^2)$ freezes in the infrared.
On the other hand, at large $Q/\Lambda$,
$\Pit$ agrees with the standard PQCD result for $\hat{\Pi}$ up to
power corrections of \order{\Lambda^2/Q^2}. Thus
$\at(Q^2,\Lambda^2)$ equals the ordinary
$\alpha_{V}(Q^2)$ for large $Q/\Lambda$.

To see in more detail how $\at(Q^2,\Lambda^2)$ freezes
in the infrared it is useful to consider the one-loop
$\beta$-function for this coupling, 
\begin{eqnarray} 
\frac{d\, \at(Q^2,\Lambda^2)}{d \ln Q} =
-\bt(\Lambda/Q)\frac{\at^2(Q^2,\Lambda^2)}{\pi} + \cdots \, .
\end{eqnarray}
Taking the derivative of $\Pit$ with respect to $\ln{Q}$ we find
\begin{eqnarray}
\bt(\Lambda/Q)
= C_A\left[ \frac{5}{6} -2 \frac{\Lambda^2}{Q^2} +
\left(2\frac{\Lambda^3}{Q^3}
       -\frac{\Lambda}{Q}
       +\frac{1}{4}\frac{Q}{\Lambda}\right)
\ln\frac{2\Lambda+Q}{|2\Lambda-Q|} \right]
\end{eqnarray}
which is plotted in Fig.~\ref{fig:beta}. The figure shows that
the running of the coupling has essentially ceased for
$Q \lsim \Lambda$.
From this it follows that if $\nu^2$ is small compared to $\Lambda^2$
then the coupling freezes in the infrared
before the pole at $Q^2=\nu^2$ is reached. The figure also illustrates
the logarithmic singularity of $\bt$ at $Q=2\Lambda$.

\FIGURE[t]{\epsfig{figure=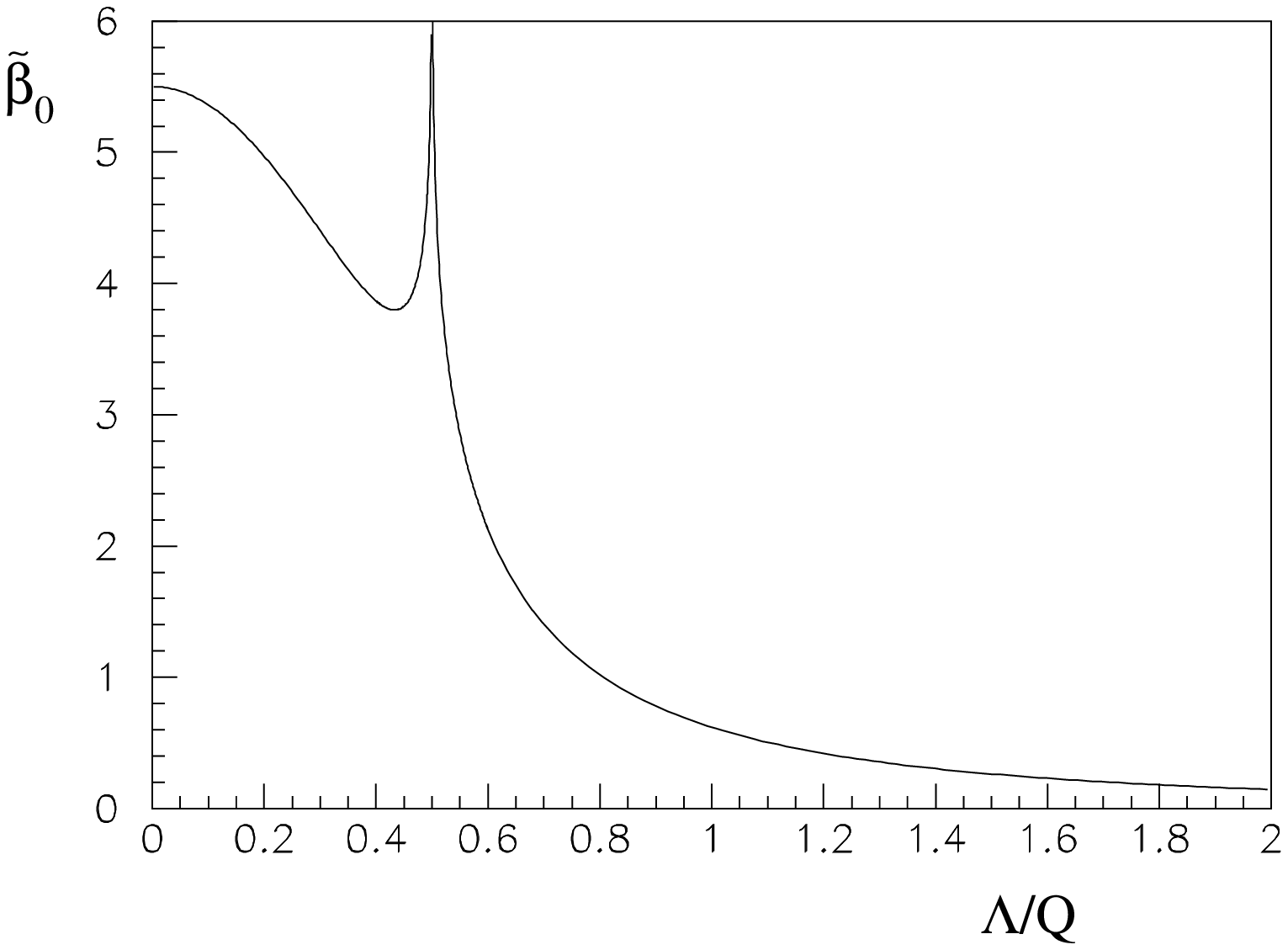,width=10cm}
\caption{ The one-loop coefficient $\bt(\Lambda/Q)$ of
the $\beta$-function for the modified running
coupling $\at(Q^2,\Lambda^2)$.}
\label{fig:beta}}

\section{Summary and conclusions}

We have explored the freedom to modify the
boundary conditions of the perturbative expansion in QCD.
More precisely we considered a specific modification, called
Perturbative Gluon Condensate Dynamics or PGCD,
where a low-energy ``sea'' of
gluons is added to the asymptotic states by modifying
the $i\varepsilon$ prescription for gluon (and ghost)
propagators. As a consequence the gluon degrees of freedom freeze
below a scale $\Lambda$,
analogously to the behaviour of fermions in a Landau liquid.
The gluon sea will scatter high-energy quarks and gluons,
preventing them from forming free asymptotic states.

In order to investigate the physical relevance of the
PGCD expansion we calculated the one-loop correction $\hat{\Pi}$
to the QCD potential between a static quark - antiquark pair.
For large $Q^2$ we found that
$\hat{\Pi}$ is unchanged up to power-corrections of
\order{\Lambda^2/Q^2}. Thus the
short distance structure of PGCD agrees with standard PQCD.
At small $Q^2$, on the other hand,
we found infrared-sensitive contributions to $\hat{\Pi}$ of
\order{\nu^2/Q^2} which after resummation give the gluon a tachyonic mass
$m_{g,\mbox{\tiny eff}}^2=-\nu^2$.
The remaining part of $\hat{\Pi}$ is constant
in the limit $Q/\Lambda \to 0$ and gives an effective
coupling $\at(Q^2,\Lambda^2)$ which freezes for $Q \lsim \Lambda$.

Our result may be summarized by the expression for the static potential
\beq \label{pgcdpot}
V(Q^2) = -4\pi C_F\frac{\at(Q^2,\Lambda^2)}{Q^2-\nu^2} =
   -4\pi C_F\frac{\at}{Q^2}\left(1+\frac{\nu^2}{Q^2}+\ldots \right) \, .
\eeq
By comparison we recall that at a finite {\it quark} density,
described by modifying the
$i\varepsilon$ prescription of the quark propagator, Debye
screening generates a {\it positive} gluon mass squared.
In coordinate space the $\nu^2/Q^2$ correction term in (\ref{pgcdpot})
corresponds to a linear confining potential.
The physical interpretation of our results for
$Q^2\lsim \nu^2$ requires further study.

Our renormalized one-loop correction (\ref{eq:Piren}) to the static potential
has a non-trivial dependence on $\Lambda/Q$. The fact that we obtained the same
result in two quite different gauges strongly suggests that the PGCD
prescription preserves QCD gauge invariance order by order in $\as$.
It would be desirable to prove this more generally.

\acknowledgments
We are grateful for helpful discussions with S.~J.~Brodsky, F.~Sannino and
K.~Zalewski. JR would like to thank Nordita for its hospitality
during a visit when parts of this work were done.
This work was supported in part by the EU Networks
``Hadron Physics with High Energy Electromagnetic Probes'', contract
EBR FMRX-CT96-0008 (PH), ``Electron Scattering Off Confined Partons'', contract
HPRN-CT-2000-00130 (PH) and ``Quantum Chromodynamics and the Deep Structure of
Elementary Particles'', contract FMRX-CT98-0194 (DG 12 - MIHT) (JR).


\begin{thebibliography}{999}

\bibitem{Dokshitzer}
Yu.~L.~Dokshitzer, Plenary talk at ICHEP 98, {\it Proc.
Vancouver 1998, High energy physics, Vol. 1, 305-324,} [\hepph{9812252}].

\bibitem{Hoyer}
P.~Hoyer,
[\hepph{9610270}];
%
P.~Hoyer, Proc. APCTP-ICTP Conf. (Seoul, Korea, May 1997), Y. M. Cho and
M. Virasoro, Eds., World Scientific (1998), p. 148,
[\hepph{9709444}];
%
P.~Hoyer, Talk at `Workshop on Exclusive and Semi-exclusive Processes
at High Momentum Transfer' (Jefferson Lab, USA, May 1999),
World Scientific (ISBN 981-02-4355-3), p. 3,
[\hepph{9908501}].

\bibitem{NN}
H.~B.~Nielsen and M.~Ninomiya,
[\hepth{9808108}].

\bibitem{dirac} P. A. M. Dirac, Rev. Mod. Phys. {\bf 21} (1949) 392.

\bibitem{nrqed}
W.~E.~Caswell and G.~P.~Lepage,
\plb{167}{1986}{437}.
See also
G.~P.~Lepage,
in Proceedings of the TASI-89 Summer School, Boulder, Colorado;
T.~Kinoshita and G.~P.~Lepage,
in {\it Quantum Electrodynamics}, ed. by T.~Kinoshita, World Scientific,
Sinagapore, 1990; 
P.~Labelle,
[\hepph{9209266}].

\bibitem{KLN}
T.~Kinoshita,
{\it J.\ Math.\ Phys}.\ {\bf 3} (1962) 650;
T.~D.~Lee and M.~Nauenberg,
\pr{133}{1964}{B1549}.

\bibitem{Akhoury:1996cj}
R.~Akhoury and V.~I.~Zakharov,
\prl{76}{1996}{2238}
[\hepph{9512433}].

\bibitem{Akhoury:1997pb}
R.~Akhoury, M.~G.~Sotiropoulos and V.~I.~Zakharov,
\prd{56}{1997}{377}
[\hepph{9702270}].

\bibitem{Akhoury:1998ks}
R.~Akhoury, L.~Stodolsky and V.~I.~Zakharov,
\npb{516}{1998}{317}
[\hepph{9609368}].

\bibitem{Susskind:1976pi}
L.~Susskind,
In {\it Les Houches 1976, Proceedings,
Weak and Electromagnetic Interactions At High Energies,}
Amsterdam 1977, 207-308.

\bibitem{khriplovich}
I.~B.~Khriplovich,
{\it Sov.\ J.\ Nucl.\ Phys}.\  {\bf 10} (1970) 235.

\bibitem{Feinberg:1978rc}
F.~L.~Feinberg,
\prd{17}{1978}{2659}.

\bibitem{Fischler:1977yf}
W.~Fischler,
\npb{129}{1977}{157}.

\bibitem{Gross:1973id}
D.~J.~Gross and F.~Wilczek,
\prl{30}{1973}{1343}.


\bibitem{Politzer:1973fx}
H.~D.~Politzer,
\prl{30}{1973}{1346}.

\bibitem{Balitsky:1985iw}
I.~I.~Balitsky,
\npb{254}{1985}{166}.

\bibitem{Chetyrkin:1999yr}
K.~G.~Chetyrkin, S.~Narison and V.~I.~Zakharov,
\npb{550}{1999}{353},
[\hepph{9811275}].

\bibitem{Lavelle:1988eg}
M.~J.~Lavelle and M.~Schaden,
\plb{208}{1988}{297}.

\bibitem{Gubarev:2001eu}
F.~V.~Gubarev, L.~Stodolsky and V.~I.~Zakharov,
\prl{86}{2001}{2220}
[\hepph{0010057}].

\bibitem{Gubarev:2001nz}
F.~V.~Gubarev and V.~I.~Zakharov,
\plb{501}{2001}{28}
[\hepph{0010096}].

\bibitem{Beneke:1998rk}
M.~Beneke,
\plb{434}{1998}{115}
[\hepph{9804241}].

\bibitem{Akhoury:1998by}
R.~Akhoury and V.~I.~Zakharov,
\plb{438}{1998}{165}
[\hepph{9710487}].

\end{thebibliography}
\end{document}